\documentclass[a4paper, 11pt]{article}

\usepackage{graphicx,amsmath,amssymb,multirow,xcolor,siunitx}
\usepackage{fullpage}

\def\cond{\,\vert\,}
\newcommand{\intens}{I}

\newcommand{\meas}{{\intens}^{\text{meas}}}
\newcommand{\R}{{\mathbb R}}
\newcommand{\real}{\mathrm{Re}}

\begin{document}

\title{Parameter estimation in modelling frequency response of coupled systems using a stepwise approach} 

\author{\normalsize{P. G\"oransson${}^a$, J. Cuenca${}^b$, and T. L\"ahivaara${}^c$}} 
\date{\normalsize{${}^a$Department of Aeronautical and Vehicle Engineering, KTH Royal Institute of Technology, Sweden\\
  ${}^b$Siemens Industry Software, Belgium\\
  ${}^c$Department of Applied Physics, University of Eastern Finland, Finland}}

\maketitle

\subsection*{Abstract}

This paper studies the problem of parameter estimation in resonant,
acoustic fluid-structure interaction problems over a wide frequency
range.  Problems with multiple resonances are known to be subjected to
local minima, which represents a major challenge in the field of
parameter identification.  We propose a stepwise approach consisting
in subdividing the frequency spectrum such that the solution to a
low-frequency subproblem serves as the starting point for the
immediately higher frequency range.  In the current work, two
different inversion frameworks are used.  The first approach is a
gradient-based deterministic procedure that seeks the model parameters
by minimising a cost function in the least squares sense and the
second approach is a Bayesian inversion framework.  The latter
provides a potential way to assess the validity of the least squares
estimate. In addition, it presents several advantages by providing
invaluable information on the uncertainty and correlation between the
estimated parameters.  The methodology is illustrated on synthetic
measurements with known design variables and controlled noise levels.
The model problem is deliberately kept simple to allow for extensive
numerical experiments to be conducted in order to investigate the
nature of the local minima in full spectrum analyses and to assess the
potential of the proposed method to overcome these.  Numerical
experiments suggest that the proposed methods may present an efficient
approach to find material parameters and their uncertainty estimates
with acceptable accuracy.

\section{Introduction}

The scope of the current work is to investigate challenges arising in
the solution of optimisation problems associated with the the
performance of coupled dynamic systems or the estimation of their
model parameters.  Of particular interest is the occurrence of local
minima, often posing a major obstacle in the solution of such
problems.

There exist a vast amount of works published within the fields of
material parameter estimation and optimisation.  A comprehensive
review of the literature is beyond the scope of the present paper; the
interested reader is referred to~\cite{tarantola04,Marburg2002}.  Most
works discussing performance optimisation in structural-acoustic
systems either work on the eigenfrequencies or limit the cost function
to frequency bands containing a few resonances, see
e.g.~\cite{WADBRO2006, Lee2007, YAMAMOTO2009, JIANBIN2010,
  Alba2011561, YANG2103, Lee2015191, SHIMODA201681}.

However, the problem of interest here, i.e. estimation of model
parameters in resonant, coupled problems over a frequency range
containing a number of resonances, remains open. In particular those
cases where, over the range of frequencies studied, there is a varying
degree of interaction between the physical fields involved, are widely
agreed to be subjected to local minima.

These often renders the parameter identification questionable if at
all tractable, as demonstrated in previous works by the authors and
others, see e.g.~\cite{tanneau2006lamary, Cuenca2012,
  LindNordgren2013, Cuenca2014, niskanen17, KLAERNER201737}.  The
presence of resonances and anti-resonances in the spectrum is not the
only factor responsible for local minima.  In fact, it has been
observed in previous works, see for example ~\cite{LindNordgren2013,
  Cuenca2014}, that a sufficient degree of anisotropy is prone to
induce such phenomena as well.  The underlying cause for this is
thought to be the self-similarity of the structural response at
discrete combinations of the material, geometrical and loading
properties. However, in order to distinguish between local minimas
related to such dynamic response phenomena, and those related to the
resonant spectrum fitting in itself, robust and efficient combinations
of inversion methods are interesting to study.

The present paper addresses a method for overcoming local minima in
the estimation of material properties in coupled dynamic resonant
problems over a wide frequency band.  The proposed approach relies on
a stepwise solution, where the solution of each sub-problem is the
starting point of the next.  While a multi-step concept as such is not
new, see for example~\cite{Vanhuyse2016462} where a 3-step procedure
was employed to enhance convergence, ~\cite{Hentati201626} where
different models are employed to fit sub-sets of parameters, and
\cite{Muhumuza2018} where seismic waves are used to estimate
subsurface properties using a frequency-by-frequency inversion
approach, the proposed strategy is novel in that it specifically
addresses resonant problems by employing an incremental frequency
domain augmentation approach as discussed in the paper.

In our work, we have chosen to combine two different estimation
methods, one based on a gradient, local search
approach,~\cite{Svanberg02}, and one global search based on the
Bayesian inversion framework,~\cite{tarantola04,kaipio05,calvetti07}.
This enables us to combine the best of two analysis
frameworks,~\cite{chazot12,niskanen17}, complementing the
deterministic search with additional insight into ill-posedness,
measurement, and/or model uncertainties, etc. in the inverse problem.
This adds a major advantage as the probability density computed from
the Bayesian solution, reveals not only the different point estimates
but also tells about the correlation between different parameters and
also the credible intervals of the estimates.  In addition, while the
Monte Carlo based estimate converges to the global minimum if the
number of samples is not limited, the combination provides more
accurate initial conditions for the chain(s) and this holds the
potential of reducing the computational time.

The proposed method is here demonstrated for a simplified
one-dimensional fluid-structure problem exhibiting a complex
behaviour, chosen for its computational tractability as well as being
inexpensive to solve.  The studied problem possesses a known solution
and its fluid-structure coupling effects are limited to the interface
between the two media.  This provides the possibility to conduct
extensive numerical experiments with controlled characteristics, such
as frequency resolution, dissipative losses, nature of coupling, etc.
The model problem chosen for the current work is selected based on the
authors' previous experiences from estimation in resonant systems
involving soft materials, and the difficulties associated with local
minima~\cite{Cuenca2014,VanderKelen2015,Vanderkelen2015PT}.  In the
latter works, parameter estimation has required manual guidance to a
certain extent in order for the cost function to converge.  The
authors aim herein at assessing the convergence behaviour of the
proposed methodology, thus opening its applicability to more complex
cases.

The paper is structured as follows. The model problem formulation is
given next in Section \ref{sec:back}. The deterministic and Bayesian
inversion frameworks are introduced in Section \ref{ref:inv} while
Section \ref{sec:num} is dedicated to numerical experiments. Finally,
discussion and conclusions are given in Sections \ref{sec:dis} and
\ref{sec:con}, respectively.

\section{Background and problem formulation}\label{sec:back}

\subsection{Model problem}

\begin{figure}[!htb]
  \centering
  \includegraphics{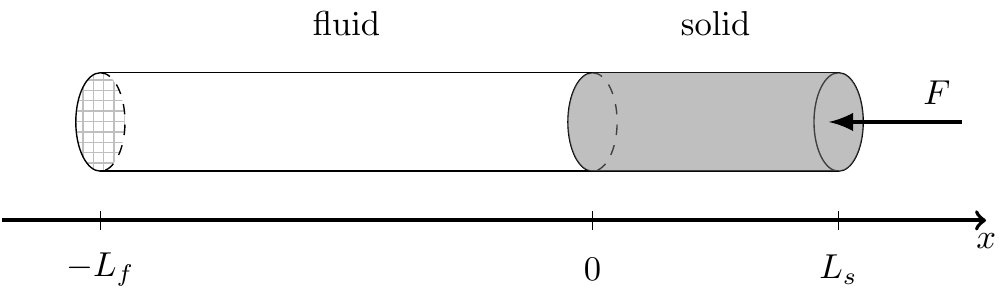}
  \caption{Schematic picture of the studied problem.}
  \label{fig:FSIex}
\end{figure}

Consider a freely moving one-dimensional (1D) solid body, excited at
one end by a harmonic force of amplitude $F$, and coupled to an air
cavity which is rigidly terminated at its other end, see
Fig.~\ref{fig:FSIex}.  Both domains are modelled in the form of the 1D
wave equation and have the same cross-section area $A$.  Let $p$
denote the acoustic pressure in the air cavity, $u$ the linear solid
displacement, $\rho_s$ the solid density, and $\rho_f$ the fluid
density.  The domain lengths are set to $L_s$ and $L_f$ for the solid
and the fluid, respectively.  The harmonic time dependence is chosen
as $e^{+j\omega t}$, where $\omega$ is the angular frequency.

The solid and the fluid are both assumed to be damped, with loss
factors $\eta_s$ and $\eta_f$, respectively.  The complex speed of
sound for the solid and the fluid may thus be written as
\begin{equation}
  c_s^2=\hat{c}_{s}^2\left( {1 + j{\eta_s}} \right)
  \quad \mbox{and} \quad 
  c_f^2=\hat{c}_{f}^2\left( {1 + j{\eta_f}} \right),
\end{equation}
where $\hat{c}_{f,s}$ denotes the amplitude of the speeds of sound.
The complex wave numbers for the fluid and solid are then
\begin{equation}
  k_{f} = \frac{\omega }{c_{f}}\quad \mbox{and} \quad k_{s} = \frac{\omega }{c_{s}}.
\end{equation}
The solution to the coupled problem is straightforward and will not be
detailed here.  The mean sound intensity level, at frequency $\omega$,
radiated from the structure to the fluid is given by
\begin{equation}
  \label{eq:PI_level}
  \intens(\theta,\xi,\omega) = 10\log\left(\frac{ |\real\left\{j\omega p_0u_0^*\right\}|}{2{\intens}_{\text{ref}}} \right),
\end{equation}
where ${\intens}_{\text{ref}}=10^{-12}\,$W$\cdot$m$^{-2}$.  The sound
intensity level $I$ is used in the current work as the target quantity
that the model should reproduce as a basis for the parameter
estimation.  In Eq.~(\ref{eq:PI_level}), the pressure and the solid
displacement (at $x=0$) are given by
\begin{eqnarray}
  \label{eq:p0}
  p_0 = p_0\left(\theta ,\xi ,\omega \right) &=& \frac{F/A}{\cos \left(k_s L_s\right) - \dfrac{\rho_s c_s}{\rho_f c_f}\tan \left(k_f L_f\right)\sin\left( k_s L_s\right)},\\
  \label{eq:u0}
  u_0 = u_0\left( \theta ,\xi ,\omega \right) &=& \frac{F/A}{\cos \left(k_s L_s\right) - \dfrac{\rho_s c_s}{\rho_f c_f}\tan \left(k_f L_f\right)\sin \left(k_s L_s\right)}\frac{\tan \left(k_f L_f\right)}{\rho_f c_f\omega}.
\end{eqnarray}
In Eqs.~(\ref{eq:PI_level}-\ref{eq:u0}), $\xi$ denotes the set of
known parameters and $\theta$ the set of unknown parameters,
\begin{eqnarray}
  \xi&=&\left\{\rho_f,\rho_s,L_f,L_s,A,F\right\},\\
  \theta&=&\left\{\hat{c}_f,\hat{c}_s,\eta_f,\eta_s\right\}.
\end{eqnarray}

\subsection{Target properties}

As stated in the introduction, the problem chosen is selected on the
basis of it being simple to model and having a low computational cost,
yet complex enough to exhibit local minima, which pose a formidable
challenge to most approaches for inverse modelling.  The fixed
parameters for the solid have been chosen to be representative of a
fairly light open-cell polymer foam, with a density low enough to
induce a strong coupling with the air cavity over a wide frequency
range.

The parameters for the target model are shown in
Table~\ref{tab:modeljust_parameters}, where $\theta^{\text{meas}}$
denotes the target parameters governing the measured system.
\begin{table}[!ht]
  \centering
  \caption{Fixed parameters $\xi$ and target parameters $\theta$ for the studied problem.} \label{tab:modeljust_parameters}
  \begin{tabular}{c|cc|cc|cc}
    &\multicolumn{2}{c}{Fluid}& \multicolumn{2}{|c|}{Solid}& \multicolumn{2}{c}{Global}\\
    \hline\hline
    \multirow{2}{*}{$\xi$} & $\rho_f$ $(\mbox{kg/m}^3)$ & $L_f$ (m) & $\rho_s$ $(\mbox{kg/m}^3)$ &$L_s$ (m) &$F$ (N) &$A$ $(\mbox{m}^2)$\\
    &1.2 & 1 & 30 & 0.338  & 1 & 0.01
    \\
    \hline
    \multirow{2}{*}{$\theta^{\text{meas}}$}&$\hat{c}_f$ (m/s) & $\hat{c}_s$ (m/s) & $\eta_f$ & $\eta_s$&\multirow{2}{*}{-}&\multirow{2}{*}{-}\\
    &343.5 & 57.735 & 0.0005 & 0.0005
  \end{tabular}
\end{table}

With the chosen target model parameters, the fundamental frequencies
and first harmonics of the uncoupled resonances of the fluid and the
solid are shown in Table~\ref{tab:modeljust_freq}.  Note that the
fundamental frequencies of the two systems are well separated, and
hence implying a moderately strong interaction between the fluid and
the structure for these.  Also, the fundamental resonance of the fluid
cavity is within 1$\,$Hz from the first harmonic of the solid, which
leads to the well-known up- and down-ward shifts in the coupled
frequencies occurring for every second resonance of the solid, as will
be shown below.
\begin{table}[!ht]
  \centering
  \caption{Fundamental $f^0_{f,s}$ and first harmonics $f^1_{f,s}$ of the uncoupled resonances of the fluid and the solid.}\label{tab:modeljust_freq}
  \begin{tabular}{cccc}
    $f_f^0 $ (Hz) &  $f_f^1 $ (Hz) & $f_s^0 $ (Hz) &  $f_s^1 $ (Hz)\\
    \hline
    171.75 & 343.5 & 85.35 & 170.7
  \end{tabular}
\end{table}

As the focus is here on the phenomena rather than computation
efficiency, a frequency range of $0.5-1000\,$Hz is used, with a
frequency resolution of $0.0153\,$Hz, in the numerical experiments
performed.

\section{Inverse problem}\label{ref:inv}

\subsection{Observation model}

Recalling the aim of this work, together with the choice of model
problem as described above, we consider also the influence of noise in
the target spectrum data, from now on referred to as measurement data.
This choice is partially made in order to avoid the "inverse crime"
trap discussed below, but also to render the measurement data set more
realistic and in this way obtain a preliminary assessment of the
robustness of the proposed method.  For the purpose of the current
work, the noise is introduced in the form of additive errors on the
sound intensity.  The observation model may thus be written as
\begin{equation}
  \label{eq:obs}
  \meas\left(\theta, \xi, \omega^{(\nu)}\right) =  {\intens}\left(\theta, \xi, \omega^{(\nu)}\right) + e,
\end{equation}
where $\omega^{(\nu)}\in \R^{N_f\times 1}$ ($N_f$ denotes the number
of frequencies) and $e$ models the measurement error.  In this work,
error $e$ is assumed to be Gaussian distributed with zero mean and
covariance $\Gamma_e = \sigma_e^2 I$, that is
$e\sim\mathcal{N}\left(0,\Gamma_e\right)$.  In the covariance matrix,
$\sigma_e$ denotes the standard deviation of the noise and $I$ is the
$N_f\times N_f$ identity matrix.

As stated in the introduction, we have chosen to apply and combine two
different approaches to the inverse problem, a gradient search method
and a method based on a Bayesian framework.  Both of these are
well-known and have been discussed in numerous papers, thus we will
not go into details about either of them as the focus here is on the
estimation and the problem of local minima in resonant systems.

Each method is run several times in order to get some statistics of
their respective convergences to the target model.  For each such
repeated computation, the starting point is generated from a uniform
distribution between the lower and the upper bounds of each of the
variables respectively.

\subsection{Numerical target}

In the current work, a purely simulation-based generation of the
measurement data is used and these are fitted with an analytical
model.  In such studies, it is of great importance to avoid committing
an "inverse crime"~\cite{kaipio05}.  Our choice for avoiding this is
to use a discretized solution, leading to a different numerical
accuracy for generating simulated measured data and the model fitted
in the inversion.  In this paper, we thus generate the measured data
set using the Comsol Multiphysics software version 5.3a while in the
inversion we use the analytic solution of the problem.

For the Comsol model, which consist of an axi-symmetric model, the
element size is $L^e_r=L^e_x=0.0058\,$m for the solid elements in a
mapped quadrilateral mesh with second order shape functions, while for
the fluid we choose elements with $L^e_x=0.0344\,$m and $L^e_r =
0.0058\,$m.  Thus the mesh is compatible along the interface between
the fluid and the solid.  For both media these element lengths observe
$L^e_{x,r} < \lambda_{\min} / 10 $ for $\lambda_{\min} =
c_s/f_{\max}$.

\begin{figure}[!ht]
  \begin{center}
    \includegraphics[width=14cm]{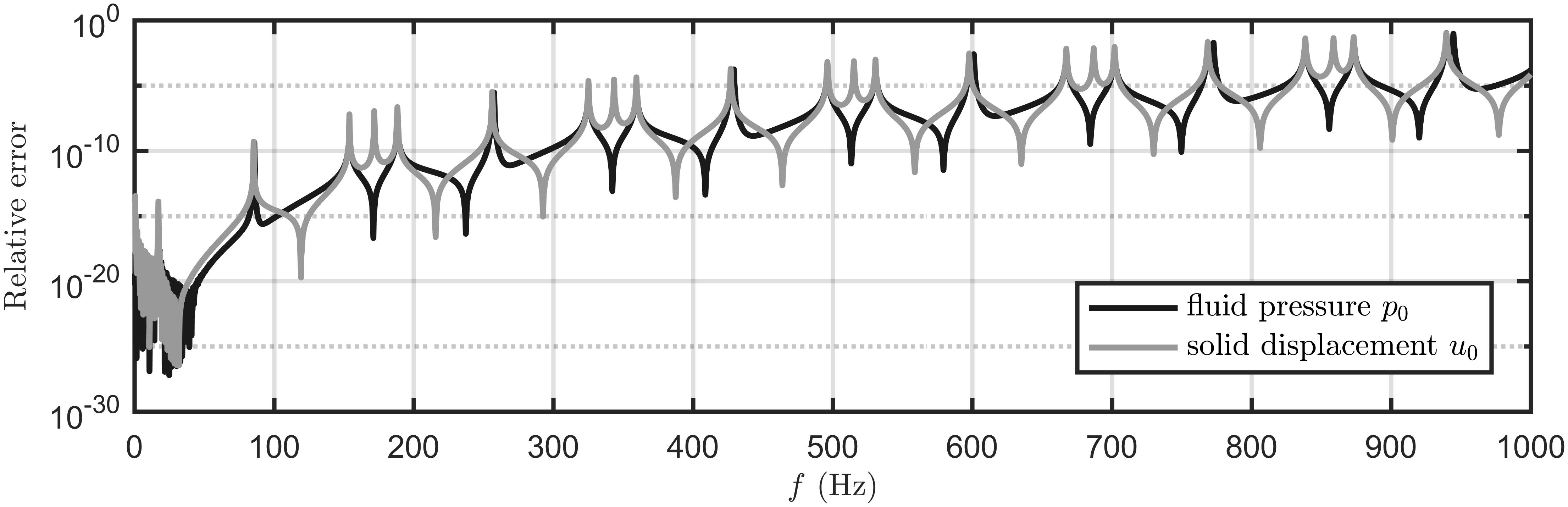}
    \caption{Relative quadratic error for the coupled Comsol and
      analytical solutions, pressure in fluid domain and displacement
      in solid.}\label{fig:Rel_error}
  \end{center}
\end{figure}

With increasing frequency, the relative quadratic error between the
discretized Comsol and the analytical solutions grows to around 7\%,
as shown in Fig.~\ref{fig:Rel_error}.

\subsection{Deterministic framework}

For the gradient search, the Globally Convergent Method of Moving
Asymptotes (GCMMA) is used in the current work~\cite{Svanberg02}.  The
cost function to be minimised is defined as
\begin{equation}
  \label{eq:costfunc}
  K(\theta) = \sum_{\ell=1}^{N_f} \frac{\left| \meas\left(\theta^{\text{meas}},\xi,\omega^{(\nu)}_{\ell}\right) -\intens\left(\theta,\xi,\omega^{(\nu)}_{\ell}\right)\right|^2}{\left| \meas \left(\theta^{\text{meas}},\xi,\omega^{(\nu)}_{\ell}\right)\right|^2},
\end{equation}
where $\theta\in\mathcal{A}$, with
$\mathcal{A}=\left[\theta_{\min},\ \theta_{\max}\right]$.  The
parameters $\theta$ are not subjected to any additional constraints.
We employ a re-scaling of all design variables such that they are
mapped to the dimensionless domain $\theta \in [1,2]$,
see~\cite{Cuenca2012} for more details.  The gradients required for
the cost function are computed using finite differencing, with a step
of $d\theta=10^{-12}$.

In the analysis, a termination criterion of the GCMMA iterations is
set to $10^{-8}$ for the variation of the cost function and $10^{-6}$
for the change in the design variables, as monitored over 5 subsequent
iterations.  As the focus is on the problem of local minima, and not
on computational efficiency, the number of iterations for each
gradient solution is allowed to reach 600 in order to ensure that the
iterations are terminated based on convergence, to facilitate the
evaluations described below.

In the following, the solutions obtained from the gradient search
method will be referred to as GCMMA.

\subsection{Bayesian framework}

In the Bayesian framework~\cite{tarantola04, kaipio05, calvetti07},
estimated variables $\theta$ are modelled as random variables.  In
this inference, the main tasks are to construct the likelihood and
prior models $\pi(\meas\cond\theta)$ and $\pi(\theta)$, respectively.
The likelihood model gives the relative probabilities that the model
generates the observations $\meas$ over all choices of parameters,
while the prior model sets constraints on the model parameters.  The
solution to the inverse problem in the Bayesian framework is coded in
the posterior distribution $\pi(\theta\cond \meas)$, which gives the
relative probability of the unknowns $\theta$ given the observations
$\meas$.  The posterior distribution is obtained by Bayes' formula
\begin{equation}
  \pi\left(\theta \cond \meas\right) \propto \pi\left(\meas \cond \theta\right)\pi(\theta).
\end{equation}

The additive observation model~(\ref{eq:obs}) allows us to write the
likelihood model as follows
\begin{equation}
  \label{eq:like}
  \pi(\meas\cond\theta) = \pi_e\left(\meas - {\intens}\left(\theta,\xi,\omega^{(\nu)}\right)\right),
\end{equation}
where $\pi_e$ is the probability density of the error $e$.  Since the
error $e$ is assumed to be Gaussian, the likelihood density can be
written as
\begin{equation}
  \pi(\meas\cond\theta) \propto \exp\left\{-\frac{1}{2}\bigg(\meas - {\intens}\left(\theta,\xi,\omega^{(\nu)}\right)\bigg)^\top\Gamma_e^{-1}\bigg(\meas - {\intens}\left(\theta,\xi,\omega^{(\nu)}\right)\bigg)\right\}.
\end{equation}
Furthermore, we use uniform prior of the form:
\begin{equation}
  \pi(\theta)\propto
  \begin{cases}{}
    1 & \text{if } \theta\in\mathcal{A}\\
    0 & \text{otherwise.}
  \end{cases}
\end{equation}

For the sampling of the posterior densities, we use the
Metropolis-Hastings
algorithm~\cite{metropolis49,metropolis53,hastings70} with an adaptive
proposal distribution scheme~\cite{Haario1999,Haario2001}.  The
adaptive algorithm automatically adjusts the spatial orientation and
size of the proposal density considering all of the target
distribution points accumulated so far, leading to a faster
convergence.

As an initial condition for the Markov chain, we use either the GCMMA
solution or random starting point and run it for 1,030,000 iterations
including a burn-in of 30,000 (unless stated otherwise in the text).
From the burn-in, the last 20,000 samples are used to provide an
initial estimate of the covariance matrix for the adaptive proposal
distribution scheme.  As the proposal density, in the first 30,000
samples, we use a zero mean Gaussian distribution, in which parameters
are assumed to be uncorrelated.

The point estimate for the parameters we use in this work is the
maximum a posteriori (MAP) which corresponds to the point in the
posterior density that has the highest probability.  As an indication
of the uncertainty in the point estimates we use the (narrowest) 95\%
credible interval, which represents the spread of the posterior
density.  In the following, the solutions obtained from the Bayesian
inversion will be referred to as MCMC (Markov chain Monte Carlo).

\section{Numerical experiments}\label{sec:num}

This section presents the results obtained for the local minima
problems in focus here.  First the inverse problem for the whole
frequency range is solved, this will be referred to as the full
spectrum estimation, including an analysis of the character of the
local minima as such.  The proposed stepwise method is then introduced
as a means to overcome these problems.

In the following examples, estimated parameters are restricted to the
interval, $\eta_s\in\ ]0,\ 1],
    \eta_f\in\ ]0,\ 1],\ c_f\in\ [10,\ 700]\ \mbox{m/s}$, and
        $c_s\in\ [10,\ 700]\ \mbox{m/s}$ which is referred to as the
        prior model in the Bayesian framework.  This choice is made in
        order to avoid adding a priori knowledge with respect to each
        of the coupled sub-systems while still keeping a reasonable
        range of possible parameter values.

\subsection{Full spectrum estimations}\label{ssec:full}

In order to explore the nature of possible local minima resulting from
GCMMA, we attempted an estimation of the parameters given in
Table~\ref{tab:modeljust_parameters}, for a noise-free target spectrum
generated through a Comsol simulation as described above.

\begin{figure}[!ht]
  \begin{center}
    \includegraphics[width=15cm]{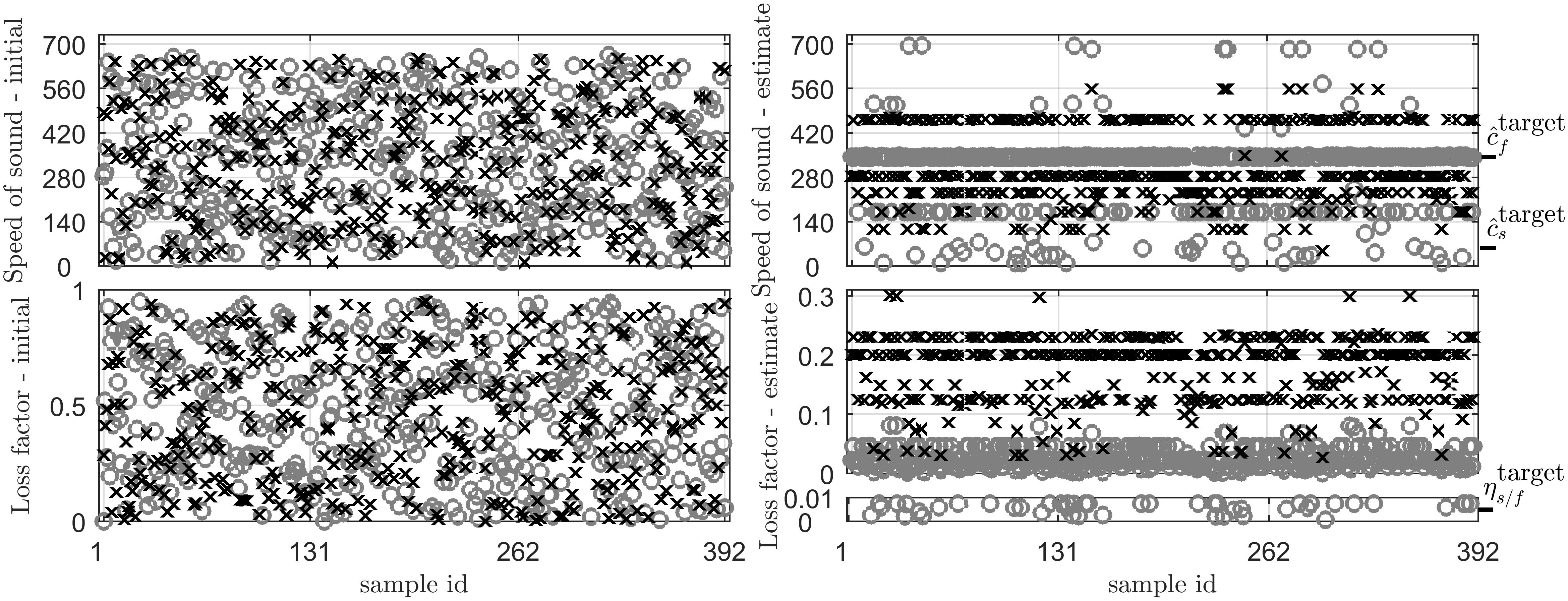}
    \caption{Left: Random starting points for full spectrum
      solution. Right: Resulting design variables, including a
      close-up on the $[0,\ 0.01]$ interval of the y-axis for the
      bottom right sub-figure. The solid parameters are denoted by
      ``$\times$'' and the fluid by
      ``${\color{gray}\circ}$''.} \label{fig:initial}
  \end{center}
\end{figure}

Figure~\ref{fig:initial} shows the initial and final parameter values
for 392 different full spectrum estimations using random starting
points, none of which converged to the global minimum.  The speed of
sound for the fluid tends to be close to the target value for most of
the solutions, while the speed of sound for the solid seems to
converge towards values that are considerably higher than the target
value.
These results illustrate the well-known difficulties encountered when
fitting a parametric model over a wide range of frequencies,
containing a large number of coupled resonances.  In fact, in order to
identify the parameters in Table~\ref{tab:modeljust_parameters} on the
full spectrum, the initial values for $\theta$ must lie in the close
neighbourhood of the target values.

\subsection{Investigating the local minima}\label{ssect:local} 

\begin{figure}[!ht]
  \begin{center}
    \includegraphics[width=14cm]{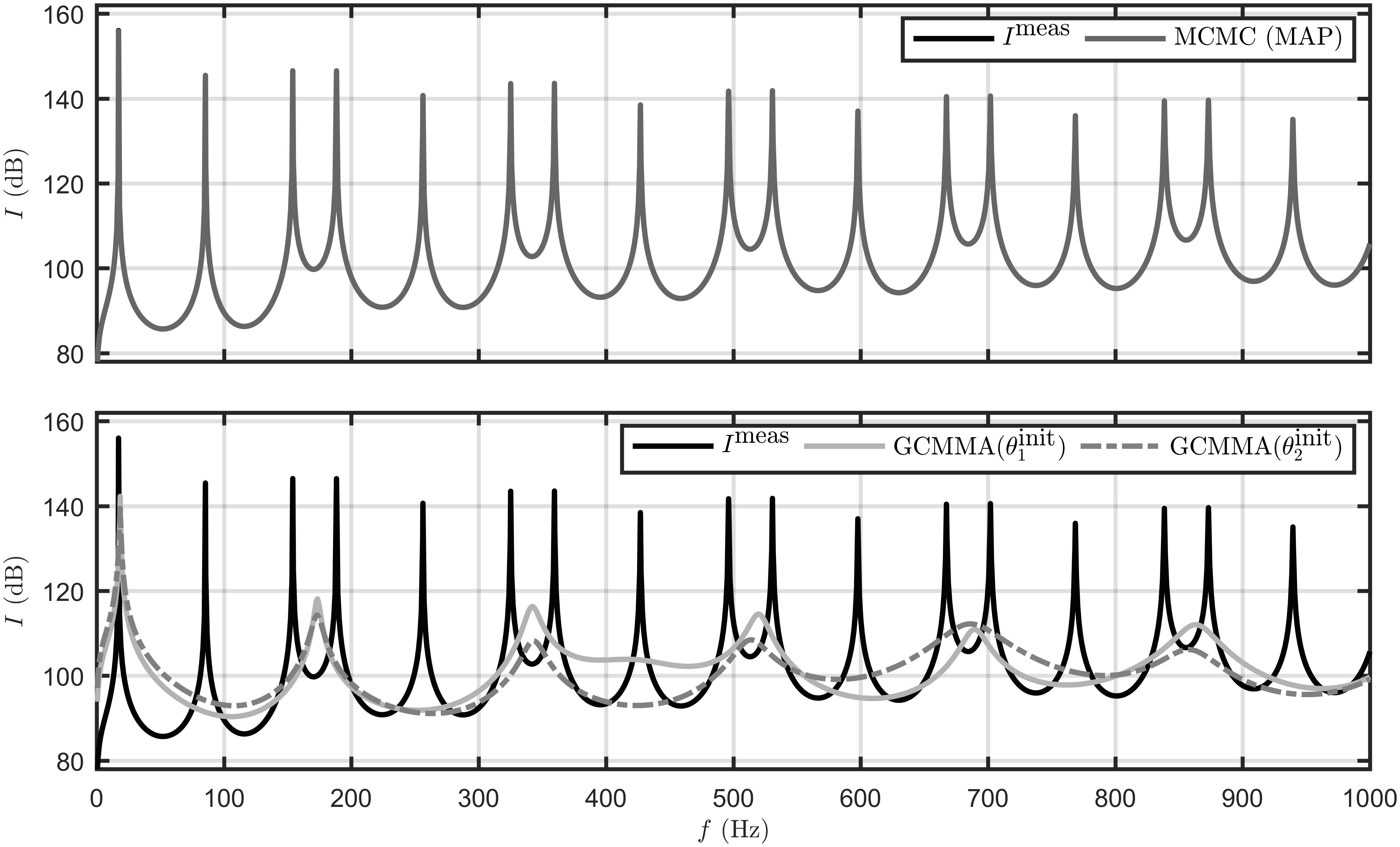}
    \caption{Fit of noise-free measurements. Top: The MCMC based MAP
      estimate after 2,000,000 samples, without use of adaptive
      proposal distribution scheme. Bottom: gradient based solutions
      $\theta^{\text{full}}_1$ and $\theta^{\text{full}}_2$ (using
      initial points $\theta^{\text{init}}_1$ and
      $\theta^{\text{init}}_2$).}\label{fig:fullspect}
  \end{center}
\end{figure}

In this section, the focus is on two randomly chosen samples from
Fig.~\ref{fig:initial}, with the intention to study the results from
the full spectrum estimations in more detail.  We have selected
starting points that are fairly close to the target value for the
fluid speed of sound, but far away from the solid, the first having
close to 10 times higher and the second nearly half the target value.
Thus, the points
\begin{flalign*}
  \theta_1^{\text{init}} &= [311.7367, 507.3036, 0.0001, 0.3023],\\ 
  \theta_2^{\text{init}} &= [324.3726, 28.8632, 0.5497, 0.4353],
\end{flalign*}
yield the following local minima,
\begin{flalign*}
  \theta_{1}^{\text{full}} &= [344.9533, 284.8150, 0.0221, 0.2007],\\
  \theta_2^{\text{full}} &= [342.8467, 462.1551, 0.0458, 0.2304].
\end{flalign*}
The results are shown in Fig.~\ref{fig:fullspect} in terms of sound
intensity level.  For both $\theta_1$ and $\theta_2$, $\hat c_f$ is
close to the value used in the target model, while for neither of the
two design points $\hat c_s$ is close to the target model value.  In
addition, as observed in the close-up $y \in [0,\ 0.01]$ in
Fig.~\ref{fig:initial}, none of the estimates for the solid loss
factor have converged to the target value, see
Table~\ref{tab:modeljust_parameters}.

The solutions obtained are quite typical for the two different
inversion approaches.  With the restrictions set, primarily in terms
of number of iterations, GCMMA terminates at different local minima
which clearly are far from the true solution, while MCMC identifies
the target model but often needs a long chain to find the global
minimum.  Note that in Fig.~\ref{fig:fullspect}, the MCMC solver does
not use adaptivity.  For the purpose of this investigation of the
local minima, the proposal density is assumed to be wide, allowing the
chain to jump between different local minima more easily.

\begin{figure}[!htb]
  \begin{center}
    \includegraphics[width=15cm]{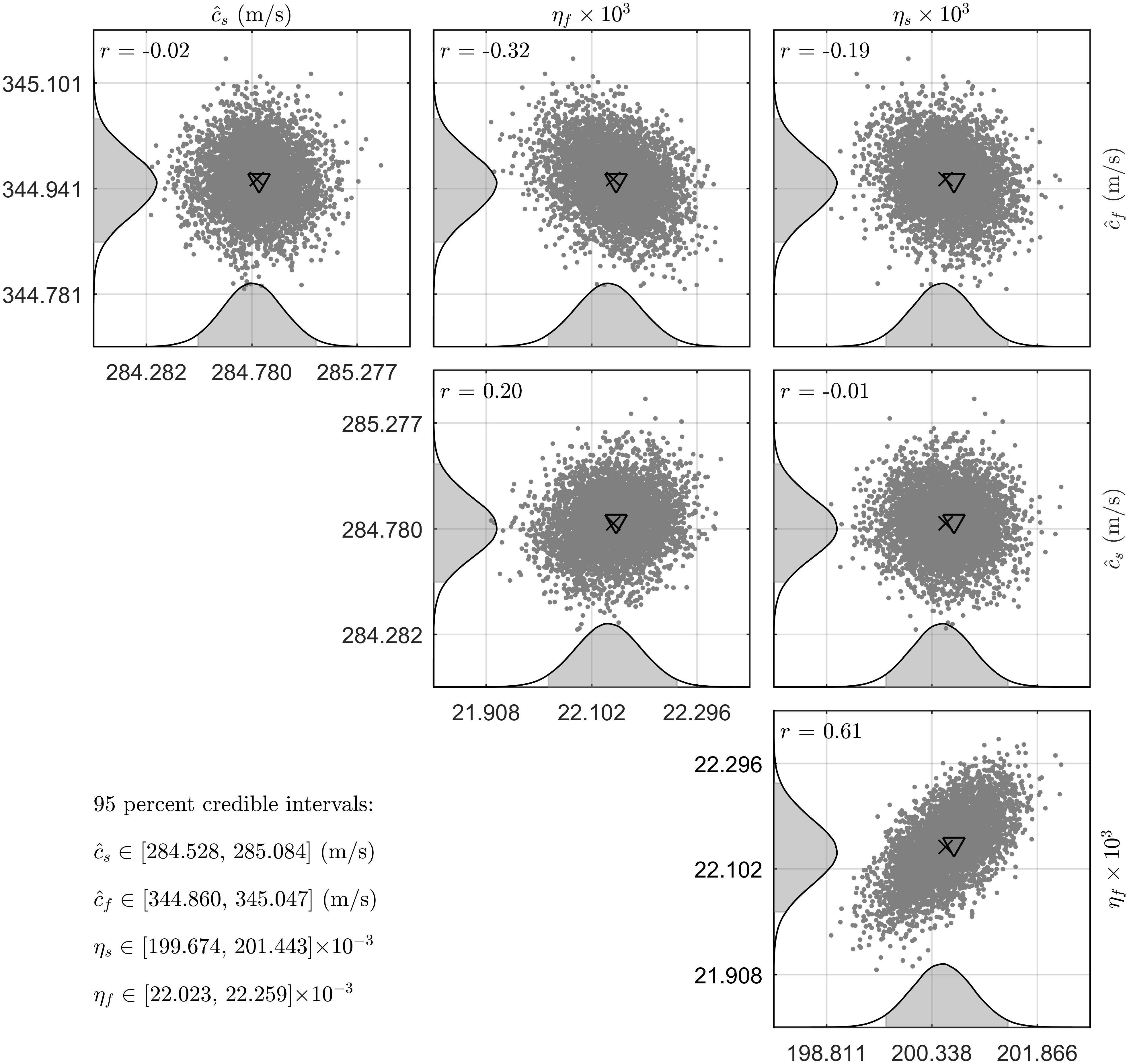}
    \caption{Two-dimensional (2D) marginal densities for the
      $\theta_1$. The cross denotes the maximum a posteriori estimate
      (MAP) and triangle is the initial condition $\theta_1$. Here the
      proposal density in the burn-in phase of the MCMC is assumed to
      be narrow in order to restrain estimated parameters to the
      neighbourhood of the GCMMA-based point estimate. The dots
      represent a sample of the MCMC points and $r$ is the Pearson
      correlation coefficient. The black curves on the bottom and left
      side of each plot are the 1D marginal posterior densities. For
      the 1D densities, the light gray zone denotes the narrowest 95\%
      credible interval (bounds given in the
      figure).}\label{fig:densities2_theta1}
  \end{center}
\end{figure}

\begin{figure}[!htb]
  \begin{center}
    \includegraphics[width=15cm]{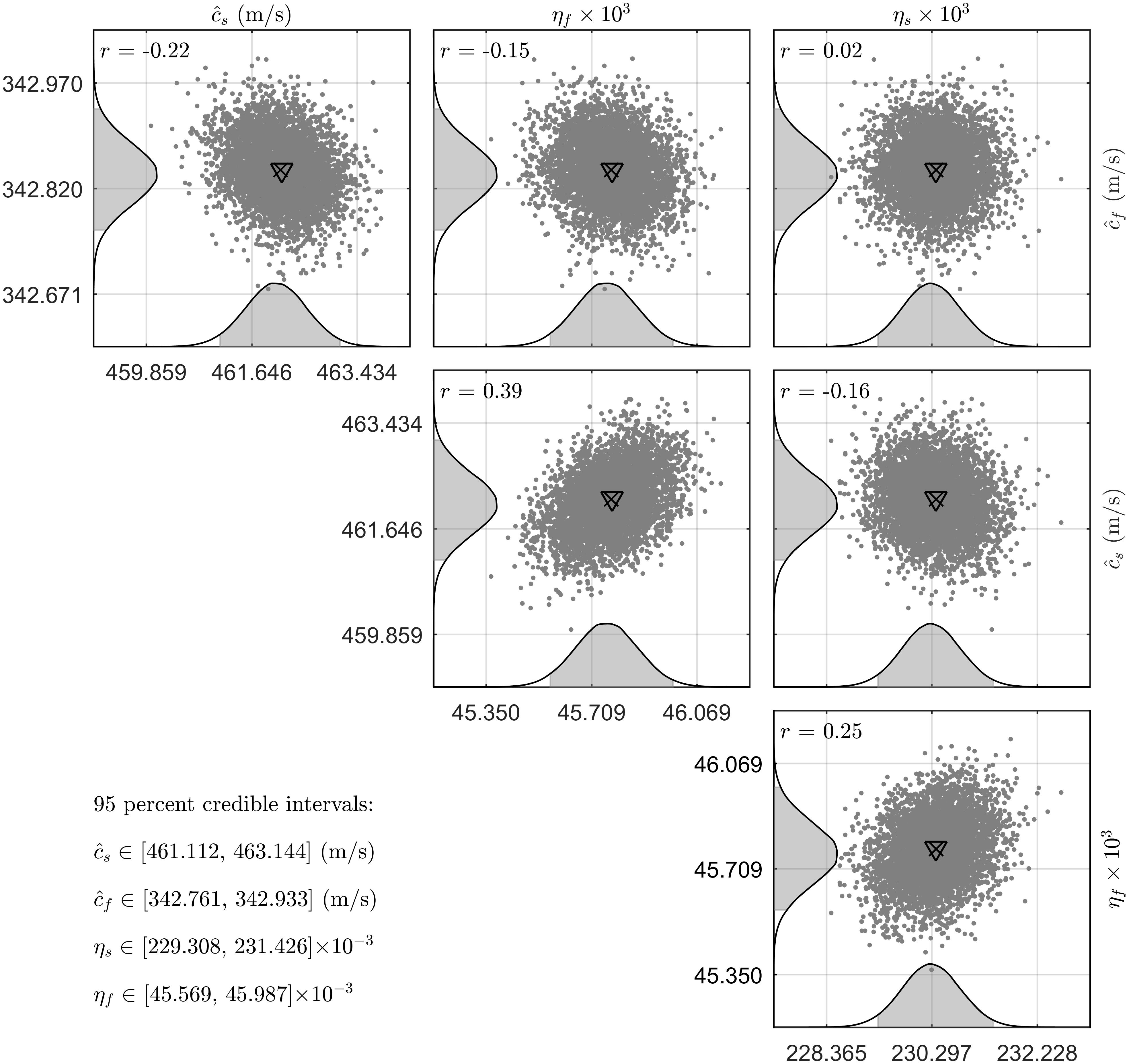}
    \caption{2D Marginal densities for the $\theta_2$, otherwise same
      caption as in
      Fig.~\ref{fig:densities2_theta1}.}\label{fig:densities2_theta2}
  \end{center}
\end{figure}

In order to confirm whether the obtained GCMMA solutions correspond to
local minima of Eq.~(\ref{eq:costfunc}), the vicinity of
$\theta_1^\text{full}$ and $\theta_2^\text{full}$ is explored using
MCMC.  Here, the MCMC algorithm uses the adaptive proposal density
after the 30,000 initial rounds but it is assumed that the proposal
density in these initial rounds is narrow in order to restrict
$\theta$ to the neighbourhood of these GCMMA solutions.
Figures~\ref{fig:densities2_theta1} and~\ref{fig:densities2_theta2}
show that both solutions obtained are indeed local minima.  All four
design variables influence the cost function to the same extent and
obviously there is no direction in which a gradient search can expect
a lowering of the cost function further.  In addition, it can be
observed from these two figures that the correlation between the model
parameters vary between almost totally uncorrelated (see e.g.~in
Fig.~\ref{fig:densities2_theta1} $\hat c_f$ and $\hat c_s$) and higher
correlation (e.g.~$\hat c_f$ and $\eta_f$; $\eta_f$ and $\eta_s$).
Furthermore, the 95\% credible intervals reveal that the local minima
are indeed narrow in all parameters.

It is worthwhile to explore the solutions obtained for the full
spectrum in some more detail.  In almost all cases we have observed,
the first resonance in the response spectrum (see
Fig.~\ref{fig:fullspect}) is well identified with respect to the
frequency but not particularly well in terms of amplitude.  The
gradient solutions appear in many cases to track the uncoupled
resonances, see for example the peak in the GCMMA solutions in
Fig.~\ref{fig:fullspect} which occurs at about 172.4 Hz and the peak
at 342.2 Hz, both very close to the uncoupled fluid resonance
frequencies given in Table~\ref{tab:modeljust_freq}.  In most of the
solutions, this appears to be achieved through converging towards
design points where, e.g.~$\hat c_f \approx \hat
c_f^{\text{target}}/2$ or $\hat c_f \approx \hat c_f^{\text{target}}$;
and $\hat c_s \approx 2, 4, 5, 8 \times \hat c_s^{\text{target}}$, see
Fig.~\ref{fig:initial}.  An interpretation of these solutions is that
they tend to combine resonances in one of the uncoupled systems with
anti-resonances in the other. Together with associated high loss
factors this leads to the fitting to the anti-resonant parts of the
target spectrum that may be observed in the lower part of
Fig.~\ref{fig:fullspect}.

As a concluding remark of this subsection, similar findings have been
observed for a case with target loss factors of 10\%, and these will
not be repeated here for the sake of conciseness.

\subsection{Stepwise approach}

A major point of attractiveness of GCMMA for solving inverse problems
is its low number of model evaluations~\cite{Svanberg02}.  However, as
illustrated above, a pattern of local minima hinders the unsupervised
determination of the global minimum or the full spectrum problem.
Observations made when studying full spectrum solutions such as those
in Sec.~\ref{ssec:full} triggered the idea for the stepwise approach
presented here for solving inverse problems involving
frequency-dependent experimental data exhibiting a resonant behaviour.
The method draws on the observations made related to the correlation
between the model parameters as well as the convergence of the
gradient method with respect to the lowest resonances of the coupled
system.  The concept is simple, yet, as detailed in what follows,
remarkably powerful and robust.

The first step of the method consists in defining a number of
sub-problems in which the problem is solved within a frequency
interval from the lowest available frequency to an incrementally
higher frequency limit, i.e.
\begin{equation}
  \label{eq:di}
  F_{\ell}^{\text{sub}} = \left[ f_{\min},\ f_{\max}^{(\ell)} \right],\quad \ell = 1,\ldots,{N^{\text{sub}}}.
\end{equation}
In the very first step, i.e.~when solving for $F_1^{\text{sub}}$,
$\theta^{\text{init}}$ is randomised such that $\hat{c}_s$ and
$\hat{c}_f$ are sampled with equal probability from the interval
$[10,\ 700]$ m/s and damping factors $\eta_s$ and $\eta_f$ from the
interval $]0,\ 1]$, i.e.~the prior model.  In each of the subsequent
    $\ell$-th steps, $\theta_\ell^{\text{init}}$ holds the parameters
    estimated from the solution obtained in the previous step,
    i.e.~$\theta_\ell^{\text{init}}=\theta_{\ell-1}$.  That is, the
    point estimate obtained from the first subdivision is used as a
    starting point for the next frequency interval until the entire
    available spectrum is covered.  The last step corresponds to an
    estimation in the full spectrum $F_{N^\text{sub}}^{\text{sub}}$
    where the starting point has been guided by the sequence of
    previous steps.

\subsection{Application to the simplified fluid-structure interaction problem}

This subsection presents the application of the proposed stepwise
approach to estimating the parameters of the target model given in
Table~\ref{tab:modeljust_parameters}.  The focus is on the convergence
of the method and therefore the computational efficiency aspects are
not discussed at this point.

The application of the proposed approach to the dataset shown in
Fig.~\ref{fig:fullspect} yields systematically satisfactory results,
only limited by the numerical stopping criteria set for the optimiser.
As these results do not convey a significant deal of information, the
performance and robustness of the method are presented on a dataset
containing additive noise.  This is done in order to demonstrate how
the stepwise approach performs for a more realistic set of data, by
including a standard deviation $\sigma_e=2\,$dB~SIL (where SIL stands
for sound intensity level) assumed for the zero mean white noise model
$e$ in Eq.~(\ref{eq:obs}).

In setting up the estimation problem, the frequency spectrum $f\in
[0.5,\ 1000]\,$Hz is divided into $N^{\text{sub}}=5$ frequency
intervals, having maximum frequencies
$f_{\max}=\{5,\ 25,\ 75,\ 150,\ 1000\}\,$Hz, see Eq.~(\ref{eq:di}).
This choice, which obviously is far from unique, is based on the
observations made in Sec.~\ref{ssect:local} on the lowest peak in the
response occurring at circa 17.5$\,$Hz and the correlations between
$\hat c_f$ and $\hat c_s$.  In particular, the first substep is aimed
at capturing the low frequency asymptotic behaviour and the parameters
governing this, in this case $\hat c_f$ and $\eta_f$ as will be
discussed later.

\begin{figure}[!ht]
  \begin{center}
    \includegraphics[width=14cm]{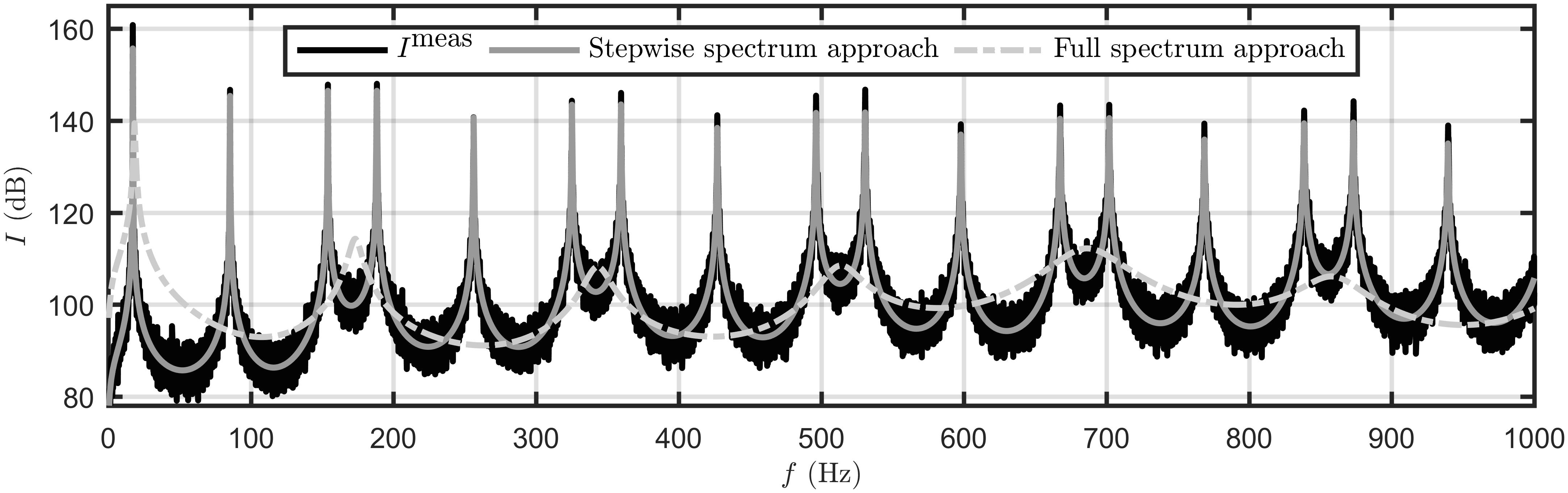}
    \caption{$\intens$ as a function of frequency: model including
      additive noise, final step in the proposed approach, and full
      spectrum estimates following
      Table~\ref{tab:estimates}.}\label{fig:estimates}
  \end{center}
\end{figure}

Using $\theta^{\text{init}} = \theta_2^{\text{init}}$, given in
Sec.~\ref{ssect:local}, as a starting point, the results in
Fig.~\ref{fig:estimates} are obtained.  The fit between the target and
the estimated model is very close, and for comparison, the less
satisfactory estimation results when using the full spectrum for the
inversion are also shown.

\begin{figure}[!ht]
  \begin{center}
    \includegraphics[width=15cm]{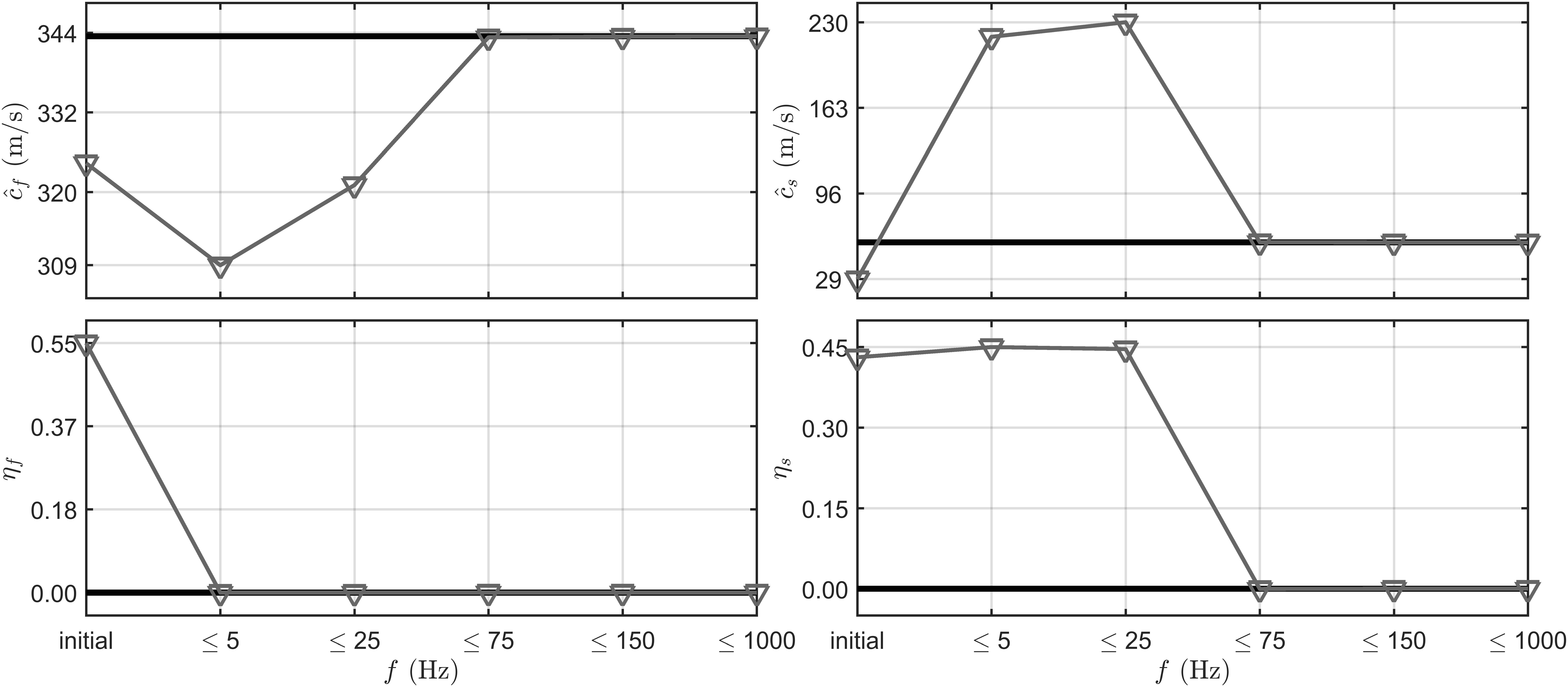}
    \caption{ Resulting estimates after each of the incremental
      frequency intervals of the stepwise approach, same data as in
      Table~\ref{tab:estimates}. Horizontal black lines show the
      target parameter values.}
    \label{fig:initial_iterations} \end{center}
\end{figure}

The convergence history for the stepwise solution is shown in
Table~\ref{tab:estimates}, together with the full spectrum estimates
and initial and target parameters. The stepwise solution and the
target values are also visualized in
Fig.~\ref{fig:initial_iterations}. Interestingly the path to the
target solution is by first finding the loss factor for the fluid
while the loss factor for the solid goes to the maximum allowed. By
the third step, all remaining parameters have been estimated. Note
that this is achieved for a quite noisy target dataset.

\begin{table}[!ht]
  \centering
  \caption{Table lists the starting point $\theta^{\text{init}}$,
    estimates for different frequency intervals of the stepwise
    approach, estimate for the full spectrum, and target
    values.}\label{tab:estimates}
  \begin{tabular}{p{25mm}|lr|S[table-format=3.4]S[table-format=3.4]S[table-format=3.4]S[table-format=3.4]}
    & $\theta$ & \multicolumn{1}{c@{}|}{$f_{\max}\,$(Hz)} & \multicolumn{1}{c@{}}{$\hat{c}_f\,$(m/s)} & \multicolumn{1}{c@{}}{$\hat{c}_s\,$(m/s)} & \multicolumn{1}{c@{}}{$\eta_f\times 1000$} & \multicolumn{1}{c@{}}{$\eta_s\times 1000$}\\
    \hline
    Starting point & $\theta^{\text{init}}$ & \multicolumn{1}{c@{}|}{-} &324.3726 & 28.8632 & 549.6625 & 435.3224\\
    \hline
    \multirow{5}{25mm}{Stepwise approach}& $\theta_1^{\text{step}}$ &5 &308.9829 & 218.6860 & 0.3969 & 454.5294\\
    &$\theta_2^{\text{step}}$&25&321.0744 & 230.1108 & 0.4735 & 450.6748\\
    &$\theta_3^{\text{step}}$&75&343.3354 & 57.8623 & 0.4985 & 0.0099\\
    &$\theta_4^{\text{step}}$&150&343.4514 & 57.7347 & 0.4956 & 0.4701\\
    &$\theta_5^{\text{step}}$&1000&343.5040 & 57.7375 & 0.4997 & 0.4999\\
    \hline
    Full spectrum &$\theta^{\text{full}}$& 1000 & 342.8564 & 460.7882 & 45.7421 & 230.8025 \\
    \hline
    Target& $\theta^{\text{meas}}$&1000 &343.5000 & 57.7350 & 0.5000 & 0.5000\\
  \end{tabular}
\end{table}

In analysing the convergence behaviour of the stepwise approach, it
should be kept in mind that in each substep the subproblems solved are
different from the full spectrum problem.  Nevertheless, the solutions
to the incremental steps provide a path of incremental partial global
minima towards the global minimum of the full spectrum problem.

\subsection{Marginal densities for the different steps}

\begin{figure}[!ht]
  \begin{center}
    \includegraphics[width=15cm]{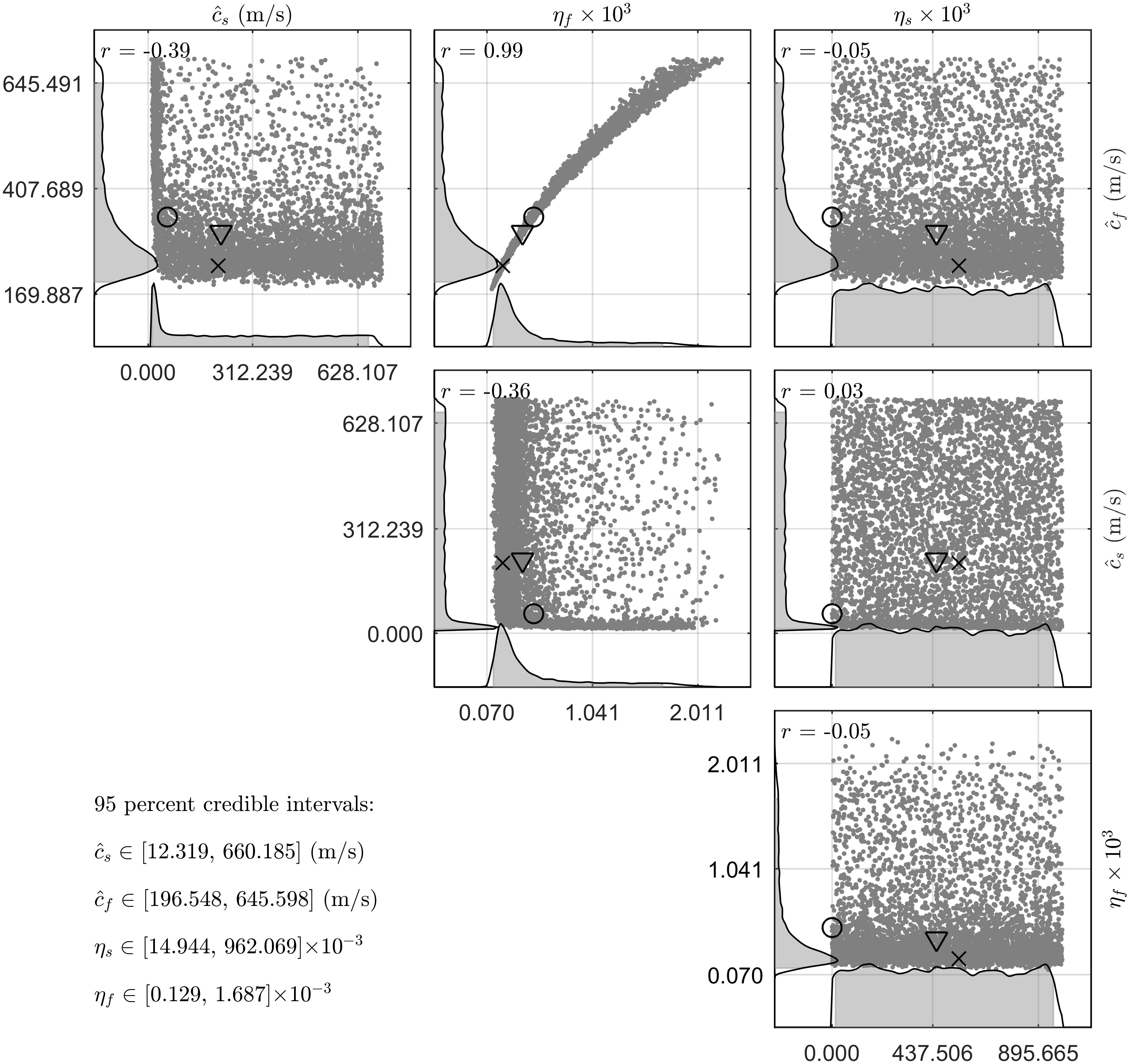}
    \caption{2D Marginal densities for the substep $f\leq\,$5 Hz
      data. The cross denotes the maximum a posteriori estimate (MAP),
      the triangle is the initial condition (based on GCMMA
      estimation, Fig.~\ref{fig:estimates}), and the circle is the
      true value. The dots represent a sample of the MCMC points and
      $r$ is the Pearson correlation coefficient. The black curves on
      the bottom and left side of each plot are the 1D marginal
      posterior densities. For the 1D densities, the light gray zone
      denotes the narrowest 95\% credible interval (bounds are given
      in the figure).}\label{fig:densities2}
  \end{center}
\end{figure}

\begin{figure}[!ht]
  \begin{center}
    \includegraphics[width=15cm]{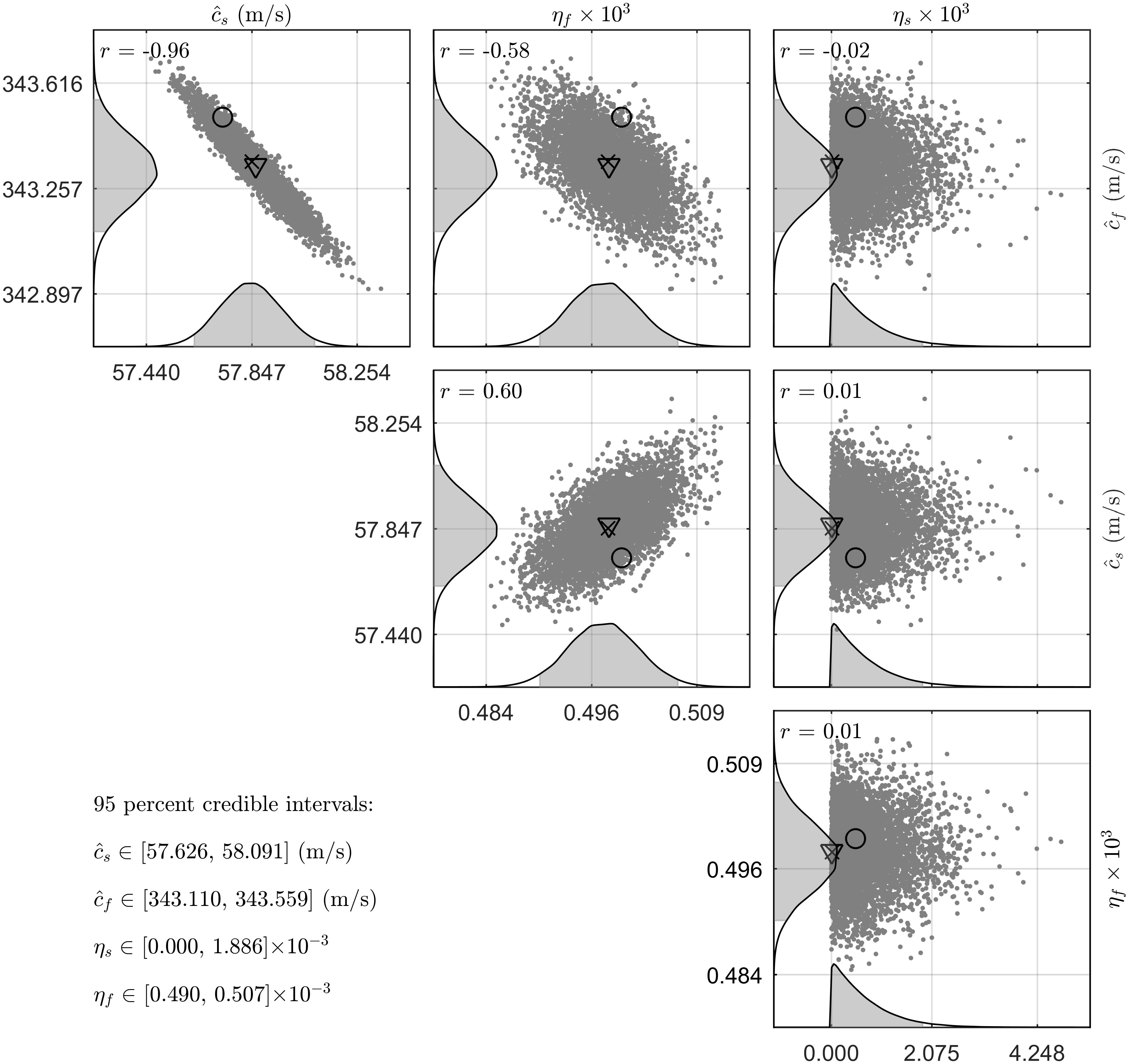}
    \caption{Results for the the substep $f\leq 75$ Hz, otherwise same
      caption as in Fig.~\ref{fig:densities2}.}
    \label{fig:densities4}
  \end{center}
\end{figure}

\begin{figure}[!ht]
  \begin{center}
    \includegraphics[width=15cm]{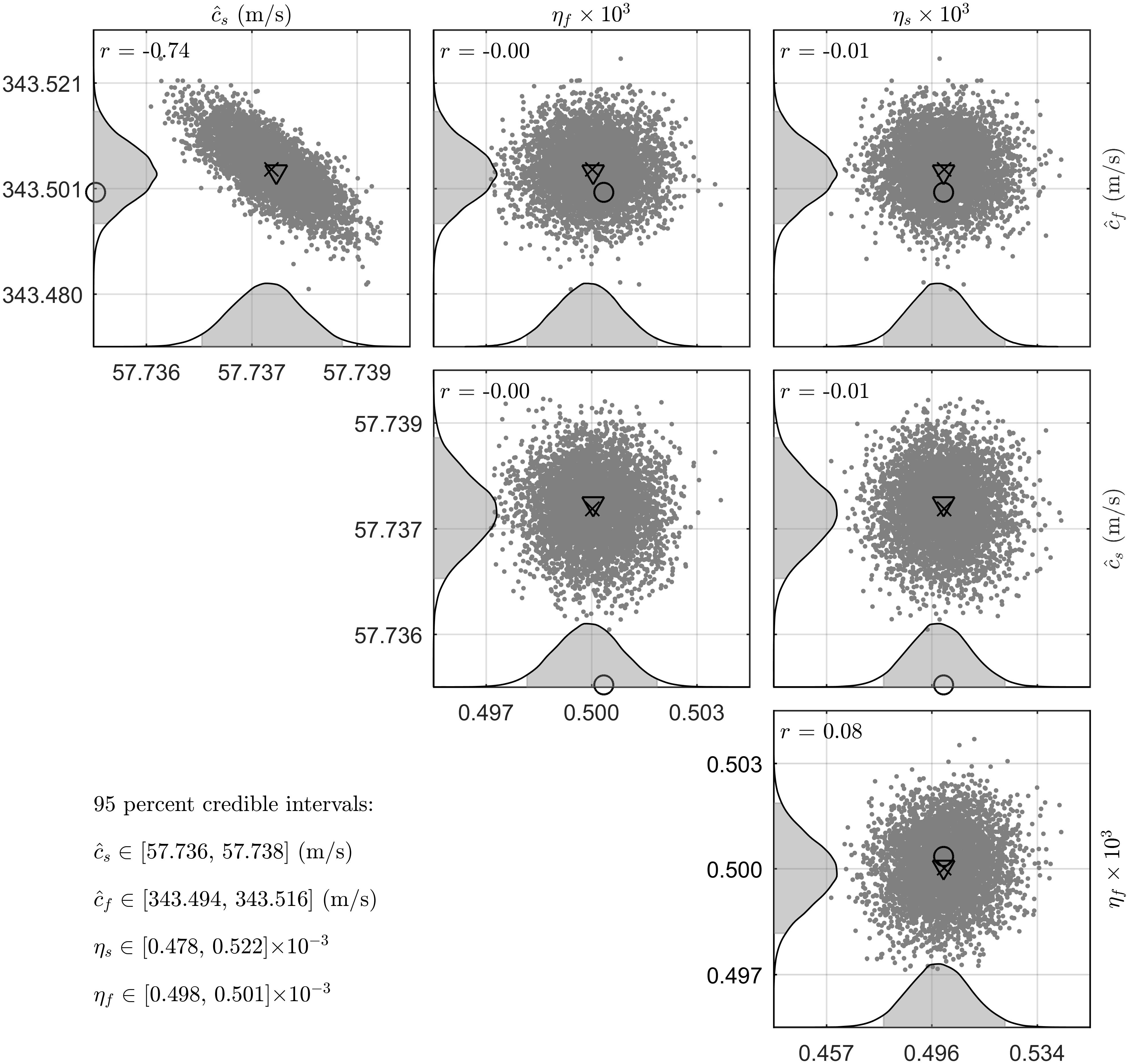}
    \caption{Results for the last substep $f\leq 1000$ Hz, otherwise
      same caption as in Fig.~\ref{fig:densities2}.}
    \label{fig:densities6}
  \end{center}
\end{figure}

Figures~\ref{fig:densities2},~\ref{fig:densities4},
and~\ref{fig:densities6} show the marginal densities at the end of
steps $f\le 5\,$Hz, $f\le 75\,$Hz, and $f\le 1000\,$Hz.  The estimates
found at the end of the first step, Fig.~\ref{fig:densities2},
indicate that there is a strong correlation between the fluid loss
factor and the fluid speed of sound, but the remaining parameters are
neither correlated to each other nor influencing the solution to the
subproblem as such.

At the 3th step, the correlation is strong between $\hat c_s$ and
$\hat c_f$, and it can be observed that both have indeed converged to
the target values.  As the full spectrum is solved in the final step,
all parameters have been estimated within 5\% relative error, with the
largest deviations occurring in the loss factors, in this case related
to the noise in the measured data set used for the estimation.  The
figures also show how the credible interval becomes narrower when the
amount of data is increased for all parameters.  The true value is
found within the narrowest 95\% credible intervals in all other cases,
except in Fig.~\ref{fig:densities6} for $\hat c_s$ but also here the
distance between the point estimate and the true value is acceptable,
given the uncertainties in the measurement data.

\begin{figure}[!ht]
  \begin{center}
    \includegraphics[width=15cm]{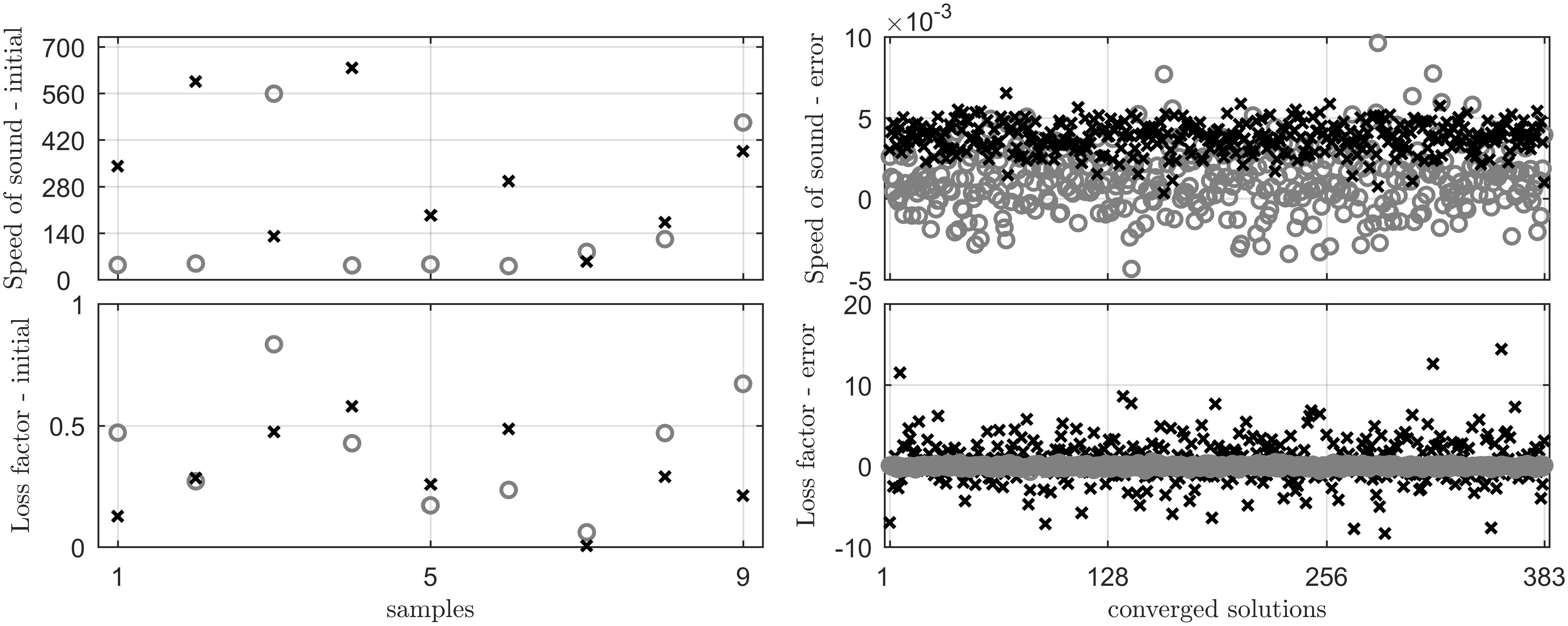}
    \caption{Left: Sample starting points that did not
      converge. Right: Relative prediction error for the converged
      samples. The solid parameters are denoted by ``$\times$'' and
      the fluid by ``${\color{gray}\circ}$''.} \label{fig:success}
  \end{center}
\end{figure}

To study the robustness of the stepwise approach when applied to the
model problem studied in the current work, we performed 392 runs,
using the same randomly chosen starting points as in
Fig.~\ref{fig:initial}.  Out of these 9 did not converge to the global
minimum, see Fig.~\ref{fig:success}.  In the converged solutions, the
speeds of sound for both fluid and solid, as well as the loss factor
for the fluid are estimated within 1\% relative prediction error,
while in particular the loss factor for the solid is less accurately
estimated.  The largest discrepancy observed for the latter is about
15\%.

\section{Discussion}\label{sec:dis}

As mentioned above, the success of the proposed stepwise approach
appears to be linked to the dominant model parameters and the
correlations between these and their variations over the different
parts of the frequency spectrum studied.  This is in agreement with
the findings in~\cite{Vanhuyse2016462,Hentati201626}.

In fact the correlation between parameters observed above for the
current model problem, may be illustrated further by taking the limit
for the Helmholtz numbers $k_f L_f,k_s L_s \ll 1$, of
Eq.~(\ref{eq:obs}).  This may be shown to be
\begin{equation}
  \label{eq:asymptote}
  \intens(\theta ,\xi ,\omega ) \approx \frac{{{{\left( {F/A} \right)}^2}}}{{\dfrac{{{\rho_f}\hat c_f^2}}{L_f}}}\dfrac{{\omega \eta_f}}{{1 - \dfrac{{{\omega ^2}}}{{\dfrac{{{\rho_s}\hat c_s^2}}{L_s}}}\left( {{\rho_s}{L_s} + 2{\rho_f}{L_f}} \right)}},\quad {\eta_f},{\eta_s} \ll 1.
\end{equation}
Accordingly, as observed above, for the first substep in the solution
of the current model problem, $\hat c_f$ and $\eta_f$ are the two
model parameters that control the behaviour of $\intens$ in
Eq.~(\ref{eq:PI_level}).  This in itself motivates our choice of the
first sub-divisions in the previous section, and in particular the
choice of the frequency range for the first substep, which in many of
the converged solutions is critical for the success.  By restricting
the first substep solution to essentially fitting
Eq.~(\ref{eq:asymptote}) to the target spectrum, the ratio between the
fluid loss factor and the square of the fluid speed of sound, which
according to Eq. \eqref{eq:asymptote} is governing the intensity at
low frequencies, is obtained, i.e.~$\sim 4.2376 10^{-09}$, already at
the end of the first step as ~$\sim 4.1573 10^{-09}$.  As a matter of
fact such ratio is invariant along the path in the marginal density
for $\hat c_f$ vs. $\eta_f$, Fig.~\ref{fig:densities2}.

A crucial aspect for the interpretation of the solution to the first
step is the fact that the marginal densities related to the solid loss
factor are nearly uniform.  This indicates that the cost function is
not sensitive to this parameter for the first step, which in turn
suggests that the sound intensity radiated by the solid into the air
cavity does not strongly depend on $\eta_s$ at very low frequencies,
as confirmed by Eq.~\eqref{eq:asymptote}.

In addition, Eq.~\eqref{eq:asymptote} and Fig.~\ref{fig:densities2}
suggest that the low-frequency asymptotic behaviour of the system
becomes increasingly sensitive to decreasing values of the speeds of
sound.  This translates into the difficulty for a gradient-based
optimiser to approach the correct solution from substantially low
values of the speeds of sound.

As identified in the convergence of repeated runs from random starting
conditions, Fig.~\ref{fig:success}, the estimates related to the solid
loss factors are the least accurate.  We relate this to the numerical
model in Comsol, used as measurement data, and the computational
errors associated with it, see Fig.~\ref{fig:Rel_error} where it is
shown that the solid displacement is the dominating source of error
for most frequencies in the spectrum.  Together with the added noise,
this renders the convergence in the final substep mostly to be
controlled by the solid loss factor, see for example
Table~\ref{tab:estimates} and Fig.~\ref{fig:densities6}.

The choice of the lower bounds of the parameter range used, in
particular for the speeds of sound, was dictated by our observations
when studying the full spectrum convergence behaviours.  In fact, the
robustness of the approach relies on the ability of the first step to
fit the asymptotic behaviour.  Lower speeds of sound would accordingly
require a lower high-frequency limit for the first step.

Although its general applicability is still to be confirmed by
exploring real parameter estimation cases, the significant improvement
in convergence experienced for the current coupled fluid-structure
problem is indisputable in our opinion.  We ran almost 400 analyses
with uniformly distributed random initial starting points, all but a
few converged to the target solution.

Even though the focus of the current work is not on computational
efficiency, it is worth noting that as the frequency range of the
first steps is relatively narrow, it is far less expensive to compute
the cost function for these as compared to the full spectrum approach.
Thus, it should be beneficial to allow for a higher number of
iterations, i.e.~employing more stringent convergence criteria in the
first increments of the stepwise estimation, in order to reduce the
efforts required in the last steps.  This is obviously a trade-off
that most likely is problem-dependent as well as depending on the
starting point and the path required to reach the global minimum.

\section{Conclusions}\label{sec:con}

Solving estimation problems over a wide frequency range with several
resonances is difficult due to local minima, which arise due to
partial fits to the spectrum.  Such an obstacle is particularly severe
for gradient search methods.  Nonetheless, also global search
approaches face difficulties, requiring a large number of iterations.

In the present paper, the proposed stepwise inversion approach is run
using several hundreds of random starting points in the parameter
design space.  The method is observed to converge to the target
solution in the vast majority of cases.  This represents a substantial
improvement of the gradient-based parameter search on its robustness
to a systematic pattern of local minima due to a resonant behaviour.
In addition, although the convergence rate depends on the initial
starting point and hence the computational cost required to obtain the
solution, the stepwise approach proposed in this paper provides
potentially a means to reduce the number of model evaluations required
to achieve convergence, by reaching increasingly good convergence as
the evaluation frequency is increased.

Our work in the current paper used a combination of MCMC and GCMMA in
order to understand the convergence behaviour.  The settings applied
for each of the two methods were partly dictated by this.
Nevertheless, in the authors' opinion, the central findings in this
paper are independent from the optimisation tool at hand.  The
stepwise method proposed herein provides a means towards circumventing
local minima in full spectrum inverse estimation problems involving
resonant structures by exploiting their asymptotic behaviour at low
frequencies.

\section*{Acknowledgements}

This work has been supported by the strategic funding of the
University of Eastern Finland and by the Academy of Finland (Finnish
Centre of Excellence of Inverse Modelling and Imaging), and the Centre
for ECO2 Vehicle Design at KTH. This article is based upon work
initiated under the support from COST Action DENORMS CA-15125, funded
by COST (European Cooperation in Science and Technology). Parts of the
computations were performed on resources provided by the Swedish
National Infrastructure for Computing (SNIC) at the Center for High
Performance Computing (PDC) at KTH.


\begin{thebibliography}{10}

\bibitem{Alba2011561}
J.~Alba, R.~Romina~Del, J.~Ramis, and J.~Arenas.
\newblock An inverse method to obtain porosity, fibre diameter and density of
  fibrous sound absorbing materials.
\newblock {\em Arch. Acous.}, 36(3):561--574, 2011.

\bibitem{calvetti07}
D.~Calvetti and E.~Somersalo.
\newblock {\em Introduction to Bayesian Scientific Computing: Ten Lectures on
  Subjective Computing (Surveys and Tutorials in the Applied Mathematical
  Sciences)}.
\newblock Springer-Verlag New York, Inc., 2007.

\bibitem{chazot12}
J.~Chazot, E.~Zhang, and J.~Antoni.
\newblock Acoustical and mechanical characterization of poroelastic materials
  using a {B}ayesian approach.
\newblock {\em J. Acoust. Soc. Am.}, 131(6):4584--4595, 2012.

\bibitem{Cuenca2012}
J.~Cuenca and P.~G\"{o}ransson.
\newblock Inverse estimation of the elastic and anelastic properties of the
  porous frame of anisotropic open-cell foams.
\newblock {\em J. Acoust. Soc. Am.}, 132(2):621, 2012.

\bibitem{Cuenca2014}
J.~Cuenca, C.~{Van der Kelen}, and P.~G\"{o}ransson.
\newblock A general methodology for inverse estimation of the elastic and
  anelastic properties of anisotropic open-cell porous materials with
  application to a melamine foam.
\newblock {\em J. Appl. Phys.}, 115(8):084904, Feb. 2014.

\bibitem{JIANBIN2010}
J.~Du and N.~Olhoff.
\newblock {Topological design of vibrating structures with respect to optimum
  sound pressure characteristics in a surrounding acoustic medium}.
\newblock {\em {Struct. Multidiscip. O.}}, {42}({1}):{43--54}, {2010}.

\bibitem{Haario1999}
H.~Haario, E.~Saksman, and J.~Tamminen.
\newblock Adaptive proposal distribution for random walk {M}etropolis
  algorithm.
\newblock {\em Comput. Stat.}, 14:375--396, 1999.

\bibitem{Haario2001}
H.~Haario, E.~Saksman, and J.~Tamminen.
\newblock An adaptive {M}etropolis algorithm.
\newblock {\em Bernoulli}, 7:223--242, 2001.

\bibitem{hastings70}
W.~Hastings.
\newblock Monte {C}arlo sampling methods using {M}arkov chains and their
  applications.
\newblock {\em Biometrika}, 57(1):97--109, 1970.

\bibitem{Hentati201626}
T.~Hentati, L.~Bouazizi, M.~Taktak, H.~Trabelsi, and M.~Haddar.
\newblock Multi-levels inverse identification of physical parameters of porous
  materials.
\newblock {\em Appl. Acoust.}, 108:26--30, 2016.

\bibitem{kaipio05}
J.~Kaipio and E.~Somersalo.
\newblock {\em Statistical and Computational Inverse Problems}.
\newblock Springer-Verlag, 2005.

\bibitem{KLAERNER201737}
M.~Klaerner, M.~Wuehrl, L.~Kroll, and S.~Marburg.
\newblock {FEA}-based methods for optimising structure-borne sound radiation.
\newblock {\em Mech. Syst. Signal Pr.}, 89:37--47, 2017.

\bibitem{Lee2015191}
J.~S. Lee, P.~G{\"o}ransson, and Y.~Y. Kim.
\newblock Topology optimization for three-phase materials distribution in a
  dissipative expansion chamber by unified multiphase modeling approach.
\newblock {\em Comput. Methods Appl. Mech. Engrg.}, 287:191--211, 2015.

\bibitem{Lee2007}
J.~S. Lee, E.~I. Kim, Y.~Y. Kim, J.~S. Kim, and Y.~J. Kang.
\newblock Optimal poroelastic layer sequencing for sound transmission loss
  maximization by topology optimization method.
\newblock {\em J. Acoust. Soc. Am.}, 122(4):2097--2106, 2007.

\bibitem{LindNordgren2013}
E.~Lind~Nordgren, P.~G\"{o}ransson, J.-F. De\"{u}, and O.~Dazel.
\newblock Vibroacoustic response sensitivity due to relative alignment of two
  anisotropic poro-elastic layers.
\newblock {\em J. Acoust. Soc. Am.}, 133(5):EL426--30, 2013.

\bibitem{Marburg2002}
S.~Marburg.
\newblock {Developments in structural-acoustic optimization for passive noise
  control}.
\newblock {\em {Arch. Comput. Method. E.}}, {9}({4}):{291--370}, {2002}.

\bibitem{metropolis53}
N.~Metropolis, A.~Rosenbluth, M.~Rosenbluth, A.~Teller, and E.~Teller.
\newblock Equation of state calculations by fast computing machines.
\newblock {\em J. Chem. Phys.}, 21(6):1087--1091, 1953.

\bibitem{metropolis49}
N.~Metropolis and S.~Ulam.
\newblock The {M}onte {C}arlo method.
\newblock {\em J. Am. Stat. Assoc.}, 44(247):335--341, 1949.

\bibitem{Muhumuza2018}
K.~Muhumuza, M.~Jacobsen, T.~Luostari, and T.~L\"ahivaara.
\newblock Seismic monitoring of {CO2} injection using a distorted {B}orn
  {T}-matrix approach in acoustic approximation.
\newblock {\em J. Seism. Explor.}, 2018.
\newblock in press.

\bibitem{niskanen17}
M.~Niskanen, J.-P. Groby, A.~Duclos, O.~Dazel, J.~C.~L. Roux, N.~Poulain,
  T.~Huttunen, and T.~L\"ahivaara.
\newblock Deterministic and statistical characterization of rigid frame porous
  materials from impedance tube measurements.
\newblock {\em J. Acoust. Soc. Am.}, 142(4):2407--2418, 2017.

\bibitem{SHIMODA201681}
M.~Shimoda, K.~Shimoide, and J.-X. Shi.
\newblock Structural-acoustic optimum design of shell structures in open/closed
  space based on a free-form optimization method.
\newblock {\em J. Sound Vib.}, 366:81--97, 2016.

\bibitem{Svanberg02}
K.~Svanberg.
\newblock {A class of globally convergent optimization methods based on
  conservative convex separable approximations}.
\newblock {\em SIAM J. Optim.}, {12}:{555--573}, {2002}.

\bibitem{tanneau2006lamary}
O.~Tanneau, J.~B. Casimir, and P.~Lamary.
\newblock Optimization of multilayered panels with poroelastic components for
  an acoustical transmission objective.
\newblock {\em J. Acoust. Soc. Am.}, 120(3):1227--1238, 2006.

\bibitem{tarantola04}
A.~Tarantola.
\newblock {\em Inverse Problem Theory and Methods for Model Parameter
  Estimation}.
\newblock SIAM, 2004.

\bibitem{VanderKelen2015}
C.~Van~der Kelen, J.~Cuenca, and P.~G\"{o}ransson.
\newblock A method for characterisation of the static elastic properties of the
  porous frame of orthotropic open-cell foams.
\newblock {\em Int. J. Eng. Sci.}, 86(0):44 -- 59, 2015.

\bibitem{Vanderkelen2015PT}
C.~{Van der Kelen}, J.~Cuenca, and P.~G\"oransson.
\newblock A method for the inverse estimation of the static elastic
  compressional moduli of anisotropic poroelastic foams - with application to a
  melamine foam.
\newblock {\em Polym. Test.}, (43):123--130, 2015.

\bibitem{Vanhuyse2016462}
J.~Vanhuyse, E.~Deckers, S.~Jonckheere, B.~Pluymers, and W.~Desmet.
\newblock Global optimisation methods for poroelastic material characterisation
  using a clamped sample in a kundt tube setup.
\newblock {\em Mech. Syst. Signal Pr.}, 68-69:462--478, 2016.

\bibitem{WADBRO2006}
E.~Wadbro and M.~Berggren.
\newblock Topology optimization of an acoustic horn.
\newblock {\em Comput. Method. Appl. M.}, 196(1):420 -- 436, 2006.

\bibitem{YAMAMOTO2009}
T.~Yamamoto, S.~Maruyama, S.~Nishiwaki, and M.~Yoshimura.
\newblock {Topology design of multi-material soundproof structures including
  poroelastic media to minimize sound pressure levels}.
\newblock {\em {Comput. Method. Appl. M.}}, {198}({17-20}):{1439--1455},
  {2009}.

\bibitem{YANG2103}
R.~Yang and J.~Du.
\newblock {Microstructural topology optimization with respect to sound power
  radiation}.
\newblock {\em {Struct. Multidiscip. O.}}, {47}({2}):{191--206}, {2013}.

\end{thebibliography}

\end{document}